\documentclass[aps]{revtex4}

\usepackage{color,epsfig,graphics,amssymb,amsmath,subeqnarray,graphicx,amsthm,subfigure}
\usepackage{natbib}

\begin{document}

\title{Viscous Marangoni propulsion}
\author{Eric Lauga}
\email{elauga@ucsd.edu}
\author{Anthony M. J. Davis}
\affiliation{Department of Mechanical and Aerospace Engineering, University of California San Diego, 9500 Gilman Drive, La Jolla CA 92093-0411, USA.}
\date{\today}
 
 \begin{abstract}

Marangoni propulsion is a form of locomotion wherein an asymmetric release of surfactant by a body located at the surface of a liquid leads to its directed motion. We present in this paper a mathematical model for Marangoni propulsion in the viscous regime.  We consider the case of a thin rigid circular disk placed at the surface of a viscous fluid and whose perimeter has a prescribed concentration of an insoluble surfactant, to which the rest of its surface is impenetrable. Assuming a linearized equation of state between surface tension and surfactant concentration, we derive analytically the surfactant, velocity and pressure fields  in the asymptotic limit of low Capillary, P\'eclet and Reynolds numbers. We then exploit these results to calculate the Marangoni propulsion speed of the disk. Neglecting the stress contribution from Marangoni flows is seen to  over-predict the propulsion speed by 50\%. 
\end{abstract}
\maketitle

\section{Introduction}

The study of animal locomotion, as carried out by zoologists and organismal biologists \citep{gray68,alexander_locomotion}, has long been a source of new problems in fluid dynamics, either 
because fluid flows  pose biological constraints which deserve to be quantified \citep{vogel96}
or because biological situations lead to interesting and new fluid physics \citep{lighthill_book,childress81}.  Broadly speaking, there exist two types of self-propelled motion in fluids. In the first type, locomotion is induced by shape changes. This category includes all classical biologically-relevant work at both high and low Reynolds numbers, and encompasses flying birds \citep{alexander02} and insects \citep{maxworthy81,ellington84,alexander02,wang05}, swimming fish \citep{lighthill_book,triantafyllou00,fish06}, the locomotion of microorganisms \citep{brennen77,pedley92,fauci06,laugapowers} and interfacial propulsion \citep{bush06}.

The second class of self-propulsion is one where  motion occurs without relative movement between the body and the fluid. In that case, transport arises due to  specific interactions between the body and its surrounding environment. Beyond the  trivial case of force-driven motion such as sedimentation  \citep{sedimentation}, this category includes transport driven by  gradients, either externally applied  \citep{young59,barton89,brochard89,chaudhury92} or self-generated  \citep{paxton04,golestanian05,paxton06,howse07,ruckner07}. In this paper, we consider  a specific setup belonging to this second category, namely  locomotion at the surface of a liquid driven by self-generated surface tension gradients, a situation sometimes referred to as Marangoni propulsion \citep{bush06}. 

A gradient in surface tension at the surface of a liquid can give rise to transport for two distinct physical reasons. First, surface tension acts as a force along contact lines. A body subject to a surface tension gradient with nonzero average will thus experience a net force, and will  have to move. Well-studied examples include droplets moving on substrates with  prescribed gradients in surface energy {\citep{greenspan,brochard89,chaudhury92}}, or  reacting with a substrate to self-generate similar  asymmetries  \citep{bain94,dossantos95}.  Surface tension gradients can also lead to transport because of  Marangoni flows. The tangential stress condition at the interface between two fluids states that tangential viscous stresses are in mechanical equilibrium with gradients in surface tension, and thus any imbalance in surface tension leads to a flow  \citep{scriven60}. Examples of Marangoni transport include the motion of droplets and bubbles in thermal gradients   \citep{young59,barton89,brochard89} and droplets subject to internal releases of surfactant \citep{tsemakh04}.

\begin{figure}
\centerline{  \includegraphics[width=0.5\textwidth]{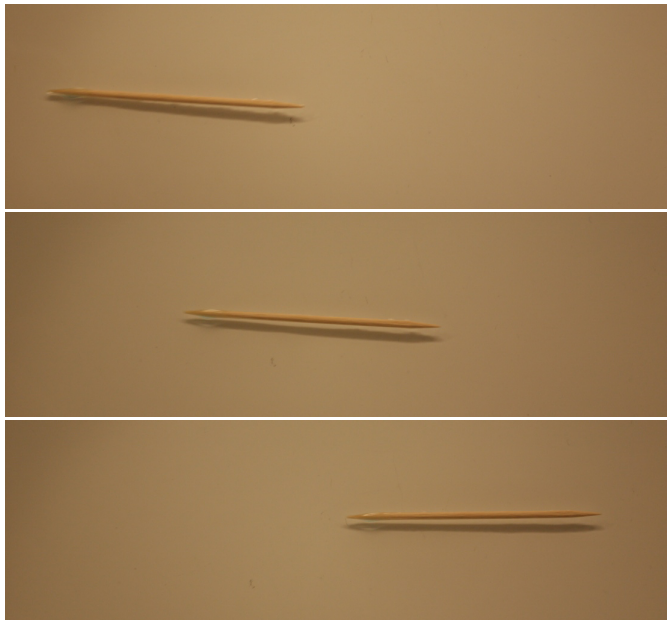}}
  \caption{Example of Marangoni propulsion in the kitchen sink. The tip of a toothpick is dipped in dishwasher liquid, and is then deposited at the free surface of a water container. The surface tension at the contact line near the rear of the toothpick is smaller than that near the front, leading to forward propulsion. The toothpick length is 6.5 cm. }
\label{kitchen}
\end{figure}

Marangoni propulsion refers to the situation where a body, located at the surface of a fluid, generates an asymmetric distribution of surface active materials, thereby prompting  a surface tension imbalance and a Marangoni flow, both of which lead to locomotion. A simple illustration of this phenomenon is illustrated in figure \ref{kitchen} using a toothpick dipped in a low surface tension dishwasher liquid. The surface tension at the contact line near the rear of the toothpick is smaller than that near the front, which leads to forward propulsion. 
Marangoni propulsion in Nature, recently reviewed by \cite{bush06}, is of relevance to the behavior of insects. The most studied  is {\it Stenus} (rove beetle), which is observed to recover from  a fall on a water surface  by quickly releasing surface active materials \citep{billard05,joy10,linsenma.ke63,schildknecht76,betz02}.  Marangoni locomotion is also used by   {\it Dianous} (rove beetle) \citep{jenkins60} and {\it Velia} (small water strider) \citep{linsenma.ke63}.

Experimentally, a lot of work has been carried out on synthetic  Marangoni propulsion, mostly regarding the dynamics of  so-called Camphor boats, where the dissolution of Camphor grains at the surface of water leads to their propulsion \citep{rayleigh,nakata97}. For that system, the Marangoni flow has been visualized  \citep{kitahata04}, and one-dimensional modeling has been proposed \citep{kohira01,kitahata04}. The combination of a chemical and hydrodynamic field has further been found to lead to interesting coupling between different bodies \citep{kohira01,soh08}, and between bodies and  boundaries \citep{nakata04}. Similar physics dictates the motion of  camphanic acid boats \citep{nakata06,nakata06_other}, phenanthroline disks \citep{nakata08,iida10}, ethanol-soaked gels on water surface \citep{bassik08} and mercury drops near oxidant crystals \citep{nakata00}.

In this paper we consider what is perhaps the simplest model of Marangoni propulsion, namely the case of a thin circular rigid disk located at the surface of a viscous fluid (\S\ref{sec:setup}).  We assume that a portion of the perimeter of the disk has a prescribed concentration of insoluble surfactant, to which the rest of the disk surface is impenetrable.  Assuming a linearized equation of state between surface tension and the concentration of surfactant, we derive mathematically in \S\ref{sec:analysis} in the limit of low Capillary, P\'eclet and Reynolds numbers  the surfactant concentration field  (\S\ref{sec: concentration}), the fluid velocity and pressure fields (\S\ref{sec:velocity}) and the propulsion speed of the disk (\S\ref{sec:propulsion}). 
Our results, and the regime in which all our assumptions are expected to be valid, are summarized in \S\ref{sec:end}. In particular, we show that neglecting the viscous stresses acting on the disk as a result of Marangoni flows over-predicts the propulsion speed by 50\%.

\section{Model problem and setup}
\label{sec:setup}

\begin{figure}
  \includegraphics[width=0.9\textwidth]{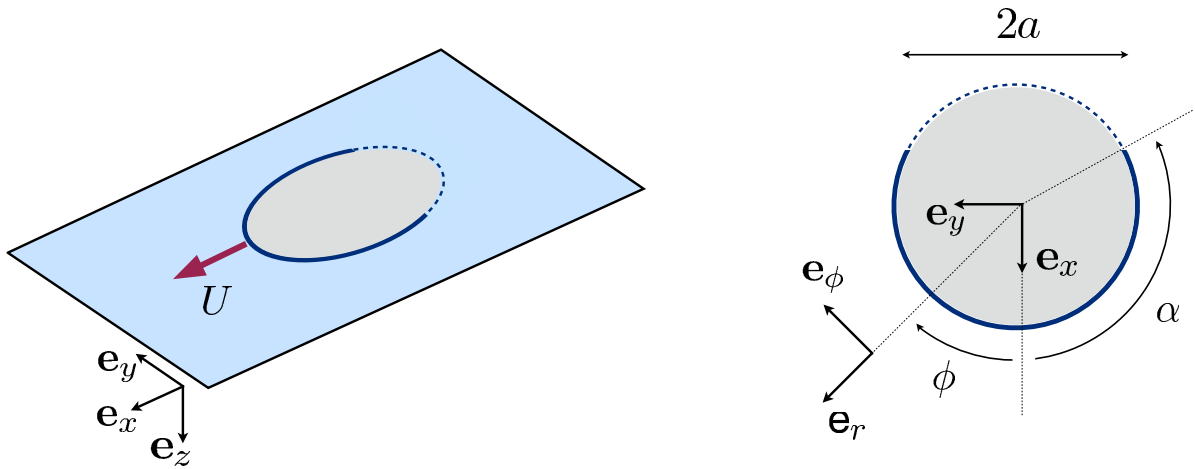}
  \caption{Setup for our model of viscous Marangoni propulsion. A solid disk  of radius $a$ is located at the surface of a semi-infinite fluid of viscosity $\mu$. The  front part of the disk ($|\phi|$ between 0 and $\alpha$ in polar coordinates) is assumed to be impermeable to surfactants. 
  The  rear part of the disk ($|\phi|$ between  $\alpha$ and $\pi$) is assumed to have a prescribed  concentration of insoluble surfactant, described by $f(|\phi|)$ in a dimensionless form, leading to forward propulsion of the disk at speed $U$.  The free surface is assumed to be undeformed by the motion of the disk, and remains flat (limit of small Capillary numbers). The P\'eclet number for the surfactant is  small enough to neglect all convective terms in the surfactant transport equation, while the Reynolds number in the fluid is small enough for the fluid motion to be described by Stokes flow. }
\label{setup}
\end{figure}

The model problem we consider in this paper is illustrated in figure \ref{setup}.  A solid disk of radius $a$ lies in the horizontal surface of otherwise  unbounded, viscous, incompressible fluid whose coefficient of viscosity
is $\mu$. Fluid motion and edgewise translation of the disk are induced
by a non-uniform surface distribution of an insoluble surfactant, of concentration $\Gamma$ per unit 
area. Specifically, the  rear part of the disk (the arc described by $|\phi|$ between  $\alpha$ and $\pi$, where $\phi$ is the polar angle in cylindrical coordinates) is assumed to have a prescribed  concentration of surfactant, described by $f(|\phi|)$ in a dimensionless form. The front part of the disk ($|\phi|$ between 0 and $\alpha$) is impenetrable to the surfactant and thus the normal derivative  of $\Gamma$ is  zero on this circular arc. {The surfactant is assumed to be insoluble in the fluid, and thus we ignore its transport to and from the bulk. For a water-air interface, this would be an appropriate approximation for surfactants such as, e.g.,  oleyl alcohol or hemicyanine \citep{handler}, and more generally organic compounds with long-chain hydrocarbon tails and polar hydrophilic heads \citep{yue}.}

{Provided we consider this problem after the initial transient in fluid flow and surfactant transport, the governing steady-state equation for surfactant concentration at the free surface \citep{stone90} is given by}
\begin {equation}\label{surf}
D\nabla_s^2 \Gamma =\nabla\cdot({\bf v}_s\Gamma), 
\end{equation}
where $D$ is the diffusion coefficient and ${\bf v}_s$ is the surface
component of the fluid velocity ${\bf v}$. This latter satisfies the
continuity equation,
\begin {equation}\label{conty}
\nabla\cdot{\bf v}=0, 
\end{equation}
and is assumed to be sufficiently small not only to allow use of the Stokes equations,
\begin {equation}\label{stokes}
\mu\nabla^2 {\bf v}=\nabla p, 
\end{equation}
where $p$ is the dynamic pressure, but also to furnish a zero P\'eclet number
approximation to the surfactant field, $\Gamma_0$, by neglecting the right hand side of
(\ref{surf}). In addition, the fluid surface is assumed to remain flat. 
In other words, we focus in this paper on the limit where all three Reynolds, P\'eclet and Capillary numbers are asymptotically small; an a posteriori derivation of the conditions under which these limits are valid is offered in  \S\ref{sec:end}.   The zero P\'eclet number surfactant field $\Gamma_0$, satisfies Laplace's equation,  and  is fully determined by its  prescribed values at the disk (\S\ref{sec: concentration}). The resulting Marangoni stresses induce the lowest approximation to ${\bf v}$ (\S\ref{sec:velocity}), which enables the translation speed of the disk to be determined (\S\ref{sec:propulsion}). We use cylindrical polar coordinates $(ar,\phi,az)$, with the origin at the center of the disk and $z$ measured vertically downwards (see figure \ref{setup}).  
In what follows, both $r$ and $z$ are thus dimensionless.

\section{Analysis}\label{sec:analysis}

\subsection{Surfactant concentration}
\label{sec: concentration}

Being a solution of the two-dimensional Laplace's equation, the
surfactant concentration, $\Gamma_0$,  in the surface $z=0$, exterior to the disk, is 
of the form
\begin {equation}\label{surf0}
\Gamma_0 a^2=c_0+2\sum_{n=1}^\infty c_n r^{-n}\cos n\phi , 
\qquad (r>1).  
\end{equation}
The dimensionless coefficients $\{c_n;n\geq 0\}$ are determined by the mixed boundary
conditions,
\begin {subeqnarray}
\Gamma_0 a^2&=&f(|\phi|) \hbox{ at } r=1,\qquad \alpha<|\phi|<\pi, \slabel{bc1}\\
\frac{\partial\Gamma_0}{\partial r}&=&0 \hbox{ at } r=1,\qquad 0<|\phi|<\alpha, \slabel{bc2}
\end{subeqnarray}
in which $f$ is a prescribed, non-constant function that is symmetric
with respect to the $\phi=\pm\pi$ direction. Application of (\ref{bc1})
and (\ref{bc2}) to (\ref{surf0}) yields a mixed boundary value problem
involving cosine series, namely,
\begin {subeqnarray}\slabel{mbv01}
c_0+2\sum_{n=1}^\infty c_n \cos n\phi &=&f(\phi) 
\qquad (\alpha<\phi\leq\pi),  \\
\sum_{n=1}^\infty nc_n \cos n\phi &=&0 \qquad (0\leq\phi<\alpha)\slabel{mbv02}.  
\end{subeqnarray}

Following the method given by \cite{sneddon} (\S5.4.3), we set
\begin {equation}\label{mbv03}
c_0+2\sum_{n=1}^\infty c_n \cos n\phi =\cos \frac{1}{2}\phi
\int_{\phi}^{\alpha}\frac{h(t)dt}{\sqrt{\cos\phi-\cos t}} ,
\qquad (0\leq\phi<\alpha),  
\end{equation}
which extends the range of (\ref{mbv01}) to the complementary interval
by using an Abel transform in anticipation of the $\sqrt{\alpha-\phi}$
behavior as $\phi\to\alpha$ from below. This immediately yields the
Fourier coefficients
\begin {equation}\label{mbv04}
c_n =\frac{1}{\pi}\left[\int_0^{\alpha}\cos\frac{1}{2}v\int_v^{\alpha}
\frac{h(t)dt}{\sqrt{\cos v-\cos t}}\cos nv\,dv +\int_{\alpha}^{\pi} 
f(v)\cos nv\,dv\right],  \qquad (n\geq 0).  
\end{equation}
When (\ref{mbv04}) is substituted into (\ref{surf0}), the summations 
can be evaluated in closed form; but first we need to relate $h(t)$ to 
$f(\phi)$.

A change in the order of integration in (\ref{mbv04}) yields
\begin {equation}\label{mbv05}
c_n =\frac{1}{\pi}\left[\int_0^{\alpha}h(t)dt\int_0^t
\frac{\cos\frac{1}{2}v\cos nv\,dv}{\sqrt{\cos v-\cos t}} 
+\int_{\alpha}^{\pi} f(v)\cos nv\,dv\right],  \qquad (n\geq 0).  
\end{equation}
{When these coefficients are substituted into the integral of
(\ref{mbv02}) between 0 and $\phi$ -- the integration constant being trivially zero -- 
the common summation can be evaluated and the double integral reduced to a single integral,
giving the Abel integral equation,}
\begin {equation}\label{mbv06}
\int_0^{\phi}\frac{h(t)\,dt}{\sqrt{\cos t-\cos\phi}} =\frac{2}{\pi}
\sin\frac{1}{2}\phi\int_{\alpha}^{\pi} \frac{f(v)\,dv}
{\cos\phi -\cos v},
  \qquad (0\leq\phi\leq\alpha).  
\end{equation}
Inversion then gives, after evaluating the $t$-integral,
\begin {equation}\label{mbv07}
h(t) =\frac{2}{\pi}\frac{d}{dt}\int_{\alpha}^{\pi} 
\frac{f(v)\sin\frac{1}{2}v}
{\sqrt{\cos t -\cos v}}dv,  \qquad (0\leq t\leq\alpha).  
\end{equation}
The first two coefficients are thus given, from (\ref{mbv05}), by
\begin {eqnarray}
\label{mbv08}
c_0&=&\frac{1}{\sqrt{2}}\int_0^{\alpha}h(t)dt +\frac{1}{\pi}
\int_{\alpha}^{\pi}f(v)\,dv=\frac{\sqrt{2}}{\pi}\int_{\alpha}^{\pi} 
\frac{f(v)\sin\frac{1}{2}v}{\sqrt{\cos\alpha -\cos v}}dv,  \\
c_1&=&\frac{1}{\sqrt{2}}\int_0^{\alpha}h(t)\cos^2\frac{1}{2}t\,dt 
+\frac{1}{\pi}\int_{\alpha}^{\pi}f(v)\cos v\,dv\nonumber\\
&=&\frac{1}{\pi\sqrt{2}}\int_{\alpha}^{\pi}\frac{f(v)\sin\frac{1}{2}v}
{\sqrt{\cos\alpha-\cos v}}(1-\cos\alpha+2\cos v)dv.  \label{mbv09}
\end{eqnarray} 
More significantly, the solution (\ref{mbv07}) enables the integral in
(\ref{mbv04}) to be expressed in terms of the prescribed $f(\phi)$ by evaluating the $t$-integral to give
\begin {equation}\label{mbv10}
\int_u^{\alpha}\frac{h(t)\,dt}{\sqrt{\cos u-\cos t}}=\frac{2}{\pi}
\int_{\alpha}^{\pi} \frac{f(v)\sin\frac{1}{2}v\,dv}{\cos u -\cos v} 
\sqrt{\frac{\cos u-\cos\alpha}{\cos\alpha-\cos v}},
 \qquad (0<u<\alpha).  
\end{equation}

Now the substitution of (\ref{mbv04}) into (\ref{surf0}) and evaluation
of the summations yields a closed form expression for the surfactant
distribution, namely,
\begin {eqnarray}\label{surf01}
\Gamma_0 a^2&=&\frac{1}{2\pi}\int_{\alpha}^{\pi}f(v)
\left[\frac{r^2-1}{r^2+1-2r\cos(\phi+v)}+
\frac{r^2-1}{r^2+1-2r\cos(\phi-v)}\right]dv \nonumber\\
&&+ \frac{1}{\pi^2}\int_0^{\alpha}\cos\frac{1}{2}u
\int_{\alpha}^{\pi}\frac{f(v)\sin\frac{1}{2}v}{\cos u -\cos v} 
\sqrt{\frac{\cos u-\cos\alpha}{\cos\alpha-\cos v}}dv \nonumber\\
&&\quad \times\left[\frac{r^2-1}{r^2+1-2r\cos(\phi+u)}
+\frac{r^2-1}{r^2+1-2r\cos(\phi-u)}\right]du, 
\end {eqnarray}
valid for $r\geq 1$. {Note that the case where  $f(\phi)=f_0$, a constant, yields a uniform surfactant concentration around the disk edge, and hence no asymmetry to generate motion.}

\subsection{Pressure and velocity field}
\label{sec:velocity}
The Marangoni stresses induce a fluid motion via the  boundary
condition for tangential stresses
\begin {equation}\label{bc3}
\mu\frac{\partial{\bf v}_s}{\partial z}=-\nabla_s\gamma ={\cal R}T_A
\nabla_s\Gamma_0, 
\end{equation}
where $\gamma$ is the surface tension, ${\cal R}$ is the gas constant, 
$T_A$ is the absolute temperature, and the linearized equation of state, 
$\gamma=\gamma_0-{\cal R}T_A \Gamma$ \citep{Gibbs,pawar96}, is invoked.

The pressure $p$ has the Fourier mode expansion,
\begin {equation}\label{press}
p=P_0(r,z)+2\sum_{n=1}^\infty P_n(r,z)\cos n\phi, 
\end{equation}
in which, since (\ref{conty}) and (\ref{stokes}) imply that $p$ is
Laplacian, the coefficients have Hankel transforms of type
\begin {equation}\label{pn}
P_n=\frac{2\mu}{a}\int_0^\infty kA_n(k)e^{-kz}J_n(kr)dk, \qquad (n\geq 0).
\end{equation}
It is then advantageous to
write the velocity field in the form
\begin{eqnarray}\nonumber
{\bf v}&=&{\bf e}_rS_0(r,z)+\sum_{n=1}^\infty 
[{\bf e}_r\cos n\phi+{\bf e}_{\phi}\sin n\phi]S_n(r,z)\\&&+\sum_{n=1}^\infty [{\bf e}_r\cos n\phi-{\bf e}_{\phi}\sin n\phi]
Q_n(r,z) +{\bf e}_z[W_0(r,z)+2\sum_{n=1}^\infty W_n(r,z)\cos n\phi].
\label{vel}\quad
\end{eqnarray} 
This ensures that the modal structure is preserved on substitution, 
with (\ref{press}), into (\ref{stokes}) which gives
\begin{eqnarray}
&&{\bf e}_r\left(\frac{\partial^2}{\partial r^2}+\frac{1}{r}
\frac{\partial}{\partial r}-\frac{1}{r^2}+
\frac{\partial^2}{\partial z^2}\right)S_0 \nonumber \\
&&+\sum_{n=1}^\infty [{\bf e}_r\cos n\phi+{\bf e}_{\phi}\sin n\phi]
\left[\frac{\partial^2}{\partial r^2}+\frac{1}{r}
\frac{\partial}{\partial r}-\frac{(n+1)^2}{r^2}+
\frac{\partial^2}{\partial z^2}\right]S_n \nonumber \\
&&+\sum_{n=1}^\infty [{\bf e}_r\cos n\phi-{\bf e}_{\phi}\sin n\phi]
\left[\frac{\partial^2}{\partial r^2}+\frac{1}{r}
\frac{\partial}{\partial r}-\frac{(n-1)^2}{r^2}+
\frac{\partial^2}{\partial z^2}\right]Q_n \nonumber \\
&&+{\bf e}_z\left[\left(\frac{\partial^2}{\partial r^2}+\frac{1}{r}
\frac{\partial}{\partial r}+\frac{\partial^2}{\partial z^2}\right)W_0
+2\sum_{n=1}^\infty \left(\frac{\partial^2}{\partial r^2}+\frac{1}{r}
\frac{\partial}{\partial r}-\frac{n^2}{r^2}+
\frac{\partial^2}{\partial z^2}\right)W_n\cos n\phi\right] \nonumber \\
&&=\frac{a}{\mu}\left[{\bf e}_r\frac{\partial P_0}{\partial r}
+\sum_{n=1}^\infty[{\bf e}_r\cos n\phi+{\bf e}_{\phi}\sin n\phi]
\left(\frac{\partial P_n}{\partial r}-\frac{n}{r}P_n\right)\right.\nonumber \\
&&\left.+\sum_{n=1}^\infty [{\bf e}_r\cos n\phi-{\bf e}_{\phi}\sin n\phi]
\left(\frac{\partial P_n}{\partial r}+\frac{n}{r}P_n\right)
+{\bf e}_z\left(\frac{\partial P_0}{\partial z}W_0+2\sum_{n=1}^\infty 
\frac{\partial P_n}{\partial z}\cos n\phi\right)\right].\label{stokesmodes}\quad \quad
\end{eqnarray}

With $P_n$ given by (\ref{pn}), it readily follows that the three sets
of partial differential equations in (\ref{stokesmodes}) have solutions of the
form
\begin {subeqnarray}
S_n&=&\int_0^\infty [D_n(k)+kzA_n(k)]e^{-kz}J_{n+1}(kr)dk, \qquad (n\geq 0),\slabel{sn} \\ 
Q_n&=&\int_0^\infty [B_n(k)-kzA_n(k)]e^{-kz}J_{n-1}(kr)dk, \qquad (n\geq 1), \slabel{qn}\\
W_n&=&\int_0^\infty kzA_n(k)e^{-kz}J_n(kr)dk, \qquad (n\geq 0),
\slabel{wn}
\end{subeqnarray}
in which the complementary function in (\ref{wn}) is chosen according
to the assumption of no normal flow in the plane $(z=0)$ of the disk 
and surfactant. Substitution of (\ref{vel}) into (\ref{conty}) gives
$$\frac{1}{r}\frac{\partial}{\partial r}(rS_0)+\frac{\partial W_0}
{\partial z}=0, $$
\begin{equation}\label{conty2}
\frac{1}{r^{n+1}}\frac{\partial}{\partial r}
(r^{n+1}S_n)+r^{n-1}\frac{\partial}{\partial r}\left(\frac{Q_n}
{r^{n-1}}\right)+2\frac{\partial W_n}{\partial z}=0, \qquad (n\geq 1),
\end{equation}
whence
\begin {equation}\label{transfms}
D_0(k)+A_0(k)=0,  \qquad D_n(k)-B_n(k)+2A_n(k)=0, \qquad (k>0,n\geq 1).
\end{equation}

These functions are determined by imposing conditions on the tangential
velocity components in the plane $z=0$. The uniform translation of the 
disk, due to the Marangoni stresses is expressed by
\begin {equation}\label{diskvel}
{\bf v}=U{\bf e}_x \,\, \hbox{ at }\,\, z=0, r<1,
\end{equation}
where $U$ is to be determined by a subsequent force balance (see \S\ref{sec:propulsion}). When 
substituted into (\ref{vel}), (\ref{diskvel}) implies that
\begin {equation}\label{diskbc1}
Q_1=U, \qquad Q_n=0\,\, (n>1), \qquad S_n=0\,\, (n\geq 0)\,\, \hbox{ at } z=0, r<1,
\end{equation}
As in \cite{davis91a, davis91b}, the Bessel functions in (\ref{sn}) and (\ref{qn}) 
are changed to $J_n$ by integration with respect to $r$, aided by the
classic recurrence relations \citep{abramowitz}. The arbitrary
constants that multiply a non-negative power of $r$ are retained and 
chosen below to eliminate square root singularities in the velocities.
Thus our required forms of the integral conditions on the interval 
$r<1$ are
\begin {subeqnarray}
\int_0^\infty D_n(k)k^{-1}J_n(kr)dk = d_n r^n,  \qquad (r<1,n\geq 0),\slabel{mbv11}\\
\int_0^\infty B_n(k)k^{-1}J_n(kr)dk=\frac{U}{2}r\delta_{1n}, 
\qquad (r<1,n\geq 1),\slabel{mbv12}
\end{subeqnarray}
where $\delta_{mn}$ denotes the Kronecker delta.

Integral conditions on the interval $r>1$ are obtained by applying the
Marangoni stress condition, (\ref{bc3}). Equation (\ref{surf0}) shows that only 
the functions $\{S_n;n\geq 0\}$ in (\ref{vel}) are directly forced by
the surface tension variations in (\ref{bc3}). Replacing $J_{n\pm 1}$ by $J_n$ as 
above, our required forms of the integral conditions become
\begin {subeqnarray}
\int_0^\infty [D_0(k)-A_0(k)]J_0(kr)dk &=&0,  \qquad (r>1), \slabel{mbv14} \\  
\int_0^\infty [D_n(k)-A_n(k)]J_n(kr)dk &=& \frac{{\cal R}T_A}{\mu a^2}c_n 
r^{-n},  \qquad (r>1,n\geq 1), \slabel{mbv13}\\
\int_0^\infty [B_n(k)+A_n(k)]J_n(kr)dk&=&b_n r^{-n} \qquad (r>1,n\geq 1). \slabel{mbv15}
\end{subeqnarray}
We now use (\ref{transfms}) to eliminate $\{A_n;n\geq 0\}$ from 
(\ref{mbv14})-(\ref{mbv15}) and obtain
\begin {subeqnarray}
\int_0^\infty D_0(k)J_0(kr)dk &=&0,  \qquad (r>1),\slabel{mbv17}   \\
\int_0^\infty D_n(k)J_n(kr)dk &=& \left[3\frac{{\cal R}T_A}{\mu a^2}c_n
+b_n\right]\frac{r^{-n}}{4},  \qquad (r>1,n\geq 1), \slabel{mbv16}\\
\int_0^\infty B_n(k)J_n(kr)dk &= &\left[\frac{{\cal R}T_A}{\mu a^2}c_n
+3b_n\right]\frac{r^{-n}}{4} , \qquad (r>1,n\geq 1).\slabel{mbv18}
\end{subeqnarray}
Equation (\ref{mbv11}) with (\ref{mbv17}), (\ref{mbv16}) and (\ref{mbv12}) with 
(\ref{mbv18}) define two infinite sets of mixed boundary value problems
of a type discussed by \cite{sneddon}  (\S4.3). The continuation of the 
integrals in (\ref{mbv11}) and (\ref{mbv12}) is given by
\begin {subeqnarray}
\int_0^\infty D_0(k)k^{-1}J_0(kr)dk &=&\frac{2d_0}{\pi} 
\int_0^{1/r}\frac{du}{\sqrt{1-u^2}} \qquad (r>1),\slabel{mbv20}   \\
\int_0^\infty D_n(k)k^{-1}J_n(kr)dk &=& \frac{2d_nr^n\Gamma(n+1)}
{\Gamma(1/2)\Gamma(n+1/2)}\int_0^{1/r}\frac{u^{2n}\,du}{\sqrt{1-u^2}}\slabel{mbv19} \\
&&+\left[3\frac{{\cal R}T_A}{\mu a^2}c_n+b_n\right]\frac{\Gamma(n-1/2)
(r^2-1)^{1/2}}{4\Gamma(1/2)\Gamma(n)r^n}  \qquad (r>1,n\geq 1), \nonumber\\
\int_0^\infty B_n(k)k^{-1}J_n(kr)dk &=& \frac{4Ur\delta_{1n}}{\pi} 
\int_0^{1/r}\frac{u^2\,du}{\sqrt{1-u^2}}\slabel{mbv21}\\
&&+\left[\frac{{\cal R}T_A}{\mu a^2}c_n+3b_n\right]\frac{\Gamma(n-1/2)
(r^2-1)^{1/2}}{4\Gamma(1/2)\Gamma(n)r^n}  \qquad (r>1,n\geq 1).\nonumber
\end{subeqnarray}

The arbitrary constants are finally determined by eliminating edge
singularities from the surface velocity components.  From (\ref{sn}) and (\ref{mbv20}), we have
\begin{eqnarray}
S_0(r,0)&=&\int_0^\infty D_0(k)J_1(kr)dk =-\frac{d}{dr}
\int_0^\infty D_0(k)k^{-1}J_0(kr)dk \nonumber \\
&=&-\frac{2d_0}{\pi}\frac{d}{dr}\arcsin\left(\frac{1}{r}\right)=0, 
\qquad (r>1), \label{s01}
\end{eqnarray}
provided that $d_0=0$. 

From  (\ref{sn}) and (\ref{mbv19}),
\begin{eqnarray}
S_n(r,0)&=&\int_0^\infty D_n(k)J_{n+1}(kr)dk =-r^n\frac{d}{dr}\left[
\frac{1}{r^n}\int_0^\infty D_n(k)k^{-1}J_n(kr)dk\right]\nonumber\\
& =&\frac{2d_n\Gamma(n+1)}{\Gamma(1/2)\Gamma(n+1/2)}\frac{r^{-(n+1)}}
{(r^2-1)^{1/2}}-\left[3\frac{{\cal R}T_A}{\mu a^2}c_n+b_n\right]
\frac{r^n\Gamma(n-1/2)}{4\Gamma(1/2)\Gamma(n)}  
\frac{d}{dr}\left[\frac{(r^2-1)^{1/2}}{r^{2n}}\right] \nonumber
\\
&=&\left[3\frac{{\cal R}T_A}{\mu a^2}c_n+b_n\right]\frac{\Gamma(n+1/2)}
{2\Gamma(1/2)\Gamma(n)}\frac{(r^2-1)^{1/2}}{r^{n+1}}  \qquad (r>1,n\geq 1), \label{sn1}
\end{eqnarray}
provided that
\begin{equation}
\frac{2d_n\Gamma(n+1)}{\Gamma(1/2)\Gamma(n+1/2)}
=\left[3\frac{{\cal R}T_A}{\mu a^2}c_n+b_n\right]\frac{\Gamma(n-1/2)}
{4\Gamma(1/2)\Gamma(n)},  \qquad (n\geq 1). 
\label{dn1}
\end{equation}

From (\ref{qn}) and (\ref{mbv21}),
\begin{eqnarray}
Q_n(r,0)&=&\int_0^\infty B_n(k)J_{n-1}(kr)dk =\frac{1}{r^n}\frac{d}{dr}
\left[r^n\int_0^\infty B_n(k)k^{-1}J_n(kr)dk\right]\nonumber \\
&=&\frac{2U}{\pi}\delta_{1n}\arcsin\left(\frac{1}{r}\right), 
\qquad (r>1,n\geq 1), \label{qn1}
\end{eqnarray}
provided that
\begin{equation}\label{bn1}
\frac{{\cal R}T_A}{\mu a^2}c_n+3b_n=\frac{8U}{\pi}\delta_{1n},  \qquad (n\geq 1). 
\end{equation}

On using (\ref{dn1}) and (\ref{bn1}) to eliminate  the coefficients $\{d_n,b_n;n\geq 1\}$,
the surface velocity is given from (\ref{vel}) by
\begin{eqnarray}
{\bf v}_s&=&\frac{2U}{\pi}\left[\arcsin\left(\frac{1}{r}\right)
({\bf e}_r\cos \phi-{\bf e}_{\phi}\sin \phi)+\frac{(r^2-1)^{1/2}}{3r^2}
({\bf e}_r\cos \phi+{\bf e}_{\phi}\sin \phi)\right]\nonumber\\
&&+\frac{4{\cal R}T_A}{3\mu a^2}\sum_{n=1}^\infty c_n 
[{\bf e}_r\cos n\phi+{\bf e}_{\phi}\sin n\phi]\frac{\Gamma(n+1/2)}
{\Gamma(1/2)\Gamma(n)}\frac{(r^2-1)^{1/2}}{r^{n+1}}, \qquad (r>1).\qquad
\label{svel}
\end{eqnarray}
Note that ${\bf v}_s=U{\bf e}_x+O[(r^2-1)^{1/2}]$ as $r\to 1+$, giving 
the expected continuous velocities. The summation in (\ref{svel}) can
be expressed as
$$(r^2-1)^{1/2}\nabla\left[-\sum_{n=1}^\infty c_n r^{-n}\cos n\phi\frac
{\Gamma(n+1/2)}{\Gamma(1/2)\Gamma(n+1)}\right], \qquad (r>1), $$
and then, by reference to $(1-x)^{-1/2}$, evaluated as in 
(\ref{surf01}).

\subsection{Marangoni propulsion speed}
\label{sec:propulsion}
The disk speed $U$ is found by enforcing that the viscous drag force,
\begin {equation}\label{drag}
2\pi\mu{\bf e}_x\int_0^1\frac{\partial Q_1}{\partial z}(r,0)r\,dr
=-2\pi\mu{\bf e}_x\int_0^\infty [B_1(k)+A_1(k)]J_1(k)dk, 
\end{equation}
is balanced by the force on the disk due to surface tension,
\begin {equation}\label{force}
{\bf e}_x\int_{-\pi}^{\pi}(\gamma)_{r=1}\cos\phi\,d\phi=
-2\pi{\bf e}_x\frac{{\cal R}T_A}{a^2}c_1,
\end{equation}
in which (\ref{vel}), (\ref{qn}), (\ref{bc3}), (\ref{surf0}) have been used (recall that $c_1$ is given explicitly by \ref{mbv09}).
Thus $U$ is determined by the equality
\begin {equation}\label{speed}
\int_0^\infty [B_1(k)+A_1(k)]J_1(k)dk=-\frac{{\cal R}T_A}{\mu a^2}c_1.
\end{equation}
The continuation of the integral in (\ref{mbv15}) is given, for $n=1$, by
\begin {eqnarray}
\int_0^\infty [B_1(k)+A_1(k)]J_1(kr)dk &=& \frac{(3U-2d_1)r}
{\pi(1-r^2)^{1/2}}+\frac{b_1}{r}\left[1-\frac{1}{(1-r^2)^{1/2}}\right]\label{speedint} \\
&=&\frac{1}{3r}\left[\frac{8U}{\pi}-\frac{{\cal R}T_A}{\mu a^2}c_1\right]
[1-(1-r^2)^{1/2}]   \qquad (r<1), \nonumber
\end{eqnarray}
and substitution into (\ref{speed}) allows us to obtain the final expression for the Marangoni propulsion speed as
\begin {equation}\label{speedU}
U=-\frac{\pi{\cal R}T_A}{4\mu a^2}c_1\cdot
\end{equation}

For the particular case, $f(\phi)=-\cos\phi$, we get, on substitution in
(\ref{mbv08}), (\ref{mbv09}),
\begin {equation}\label{c0c1}
c_0=\frac{1}{2}(1-\cos\alpha),  \qquad  c_1=-\frac{1}{8}(1+\cos\alpha)^2. 
\end{equation}
The positive surfactant concentration at the rear of the disk induces a lower
surface tension there and hence a positive speed $U$.  Note the limit 
values, $c_0(0)=0$ and $c_1(0)=-1/2$, corresponding to
$\Gamma a^2=-r^{-1}\cos\phi$ in (\ref{surf0}). Also $c_0(\pi)=1$ and 
$c_1(\pi)=0$, corresponding to a uniform surfactant of magnitude 
$f(\pi)$, which is the $\alpha\to\pi$ limit of (\ref{mbv08}). 
Similarly, the particular choice $f(\phi)=\cos 2\phi$ yields $c_1=-\frac{1}{4}(1+\cos\alpha)
\sin^2\alpha$, while $f(\phi)=-\cos 3\phi$ yields $c_1=-\frac{3}{32}
(1+\cos\alpha)(1-5\cos\alpha)\sin^2\alpha$.

\section{Discussion}\label{sec:end}

In this paper we consider a simple model setup for Marangoni propulsion: a thin rigid circular disk located at the flat free surface of a viscous fluid and releasing insoluble  surfactants with a prescribed concentration  along its edge.  As seen above, although both the geometry and setup are quite simple, the mathematical details of the derivation can be quite involved, and point to the richness of the physical problem.  {Physically, in order for the disc to move with finite speed, Eq.~\eqref{speedU} shows that the surfactant concentration cannot be uniform. From a thermodynamic point of view, the disk motion is thus an intrinsically non-equilibrium effect, where a non-uniform surfactant chemical potential   needs to be maintained along the disk edge to drive the Marangoni propulsion.}

{The steady-state} analysis in the paper makes three distinct physical assumptions, namely the limit of small Capillary, P\'eclet and Reynolds numbers. The result for the propulsion speed,  \eqref{speedU}, allows us to derive the appropriate velocity scale for the fluid, namely $U\sim{{\cal R}T_Ac_1}/{\mu a^2}$. With a natural length scale $L\sim a$, we can then use this result to  derive the regime under which  our three assumptions are expected to be correct, namely 
\begin{subeqnarray}
{\rm Ca}&=&\frac{\mu U}{\gamma_0}=\frac{{\cal R}T_Ac_1}{\gamma_0 a^2}\ll 1,\\
{\rm Pe}&=&\frac{Ua}{D}=\frac{{\cal R}T_A c_1}{\mu aD}\ll 1,\\
{\rm Re}&=&\frac{\rho Ua}{\nu}= \frac{\rho{\cal R}T_Ac_1}{\mu^2 a } \ll 1.
\end{subeqnarray}

An interesting observation from our theoretical results concerns the role of Marangoni stresses on the disk motion. An approximate method to derive \eqref{speedU} would consist in neglecting the stresses on the disk arising from Marangoni flow. In that case, $U$ would be obtained by balancing 
the direct forcing from surface tension at the contact line, $F= -2\pi {\cal R} T_A c_1/a$ (\ref{force}, with $c_1$ from \ref{mbv09}) by the viscous drag on the bottom half of the disk, $F={16}\mu a U/3$ \citep{kimbook}, leading to propulsion at the speed $ U = -3\pi {\cal R} T_A c_1/8\mu a^2$.  This result, although it has the correct sign and order of magnitude, actually over-predicts the propulsion speed \eqref{speedU} by 50\% and thus both direct surface tension forcing at the contact line and Marangoni flow play important role in determining the steady-state Marangoni propulsion speed.

\section*{Acknowledgements}
This work was supported in part by the National Science Foundation (CAREER Grant CBET-0746285 to E.L.).

\bibliography{bio_marangoni}

\end{document}